\documentclass[a4paper,12pt]{article}
\pdfoutput=1
\usepackage{epsfig,graphicx,xcolor,amsbsy,amssymb,latexsym,amsfonts,amsmath,setspace}
\usepackage{pstricks}
\usepackage{color}
\usepackage[numbers,square,comma, compress]{natbib}
\usepackage{placeins}

\usepackage{eurosym}
\usepackage[a4paper]{geometry}
\usepackage{graphicx}
\usepackage{tikz}
\usepackage{xcolor}
\usepackage{amsmath,amssymb,amsfonts,pstricks,setspace}
\usepackage{fancybox}

\numberwithin{equation}{section}

\newcommand{\bea}{\begin{eqnarray}\displaystyle}
\newcommand{\eea}{\end{eqnarray}}
\newcommand{\nn}{\nonumber \\}

\newcommand{\figref}[1]{Fig.~\protect\ref{#1}}

\title{
\vspace{-1.8cm}
\bf{Refined Topological Strings on Local $\mathbb{P}^2$}\\[15pt]}
\author{\large \textsc{Amer Iqbal\footnote{\tt amer@alum.mit.edu}~,~Can Koz\c{c}az\footnote{\tt  kozcaz@cmsa.fas.harvard.edu}}}
\date{}

\begin{document}

\maketitle

\begin{center}
\renewcommand{\thefootnote}{\fnsymbol{footnote}}\vspace{-0.5cm}
${}^{\footnotemark[2]}$ Abdus Salam School of Mathematical Sciences \\ Government College University, Lahore, PAKISTAN\\[0.4cm]
${}^{\footnotemark[2]}$ Center for Theoretical Physics, Lahore, PAKISTAN\\[0.4cm]
${}^{\footnotemark[3]}$ Center of Mathematical Sciences and Applications\\ Harvard University, Cambridge, MA 02138, USA\\[0.4cm]
\end{center}

\abstract{We calculate the refined topological string partition function of the Calabi-Yau threefold which is the total space of the canonical bundle on $\mathbb{P}^2$ (the local $\mathbb{P}^2$). The refined topological vertex formalism can not be directly applied to local $\mathbb{P}^2$ therefore we use the properties of the refined Hopf link to define a new two legged vertex which together with the refined vertex gives the partition function of the local $\mathbb{P}^2$. 
\maketitle

\section{Introduction}
The topological string theory amplitudes are useful in the study of the BPS sector of the string compactifications. ${\cal N}=2$ topological strings on Calabi-Yau threefolds capture the quantum numbers of the BPS states obtained by wrapping branes on holomorphic curves \cite{GV1,GV2}. When the Calabi-Yau threefold is non-compact, the topological string amplitudes  completely determine the prepotential, and the higher genus corrections to it, of the supersymmetric theory which lives in the spacetime \cite{Nekrasov}. Surprisingly these amplitudes also determine the superconformal index of the 5D superconformal theories which appear by M-theory compactification on non-compact Calabi-Yau threefolds \cite{Vafa:2012fi,Iqbal:2012xm}. 

Advances in the instanton computation of ${\cal N}=2$ gauge theories  motivated the refinement of the topological string amplitudes for non-compact Calabi-Yau threefolds \cite{Moore:1997dj,Nekrasov,HIV} . From the target space interpretation of the topological string theory the refinement corresponds to an explicit determination of the quantum numbers of BPS particles in M-theory compactifications on the same Calabi-Yau threefold \cite{GV1,GV2,HIV} with respect to the $SO(4)=SU(2)_{L}\times SU(2)_{R}$ part of the 5D Lorentz group. Therefore it is particularly useful to be able to calculate these refined amplitudes for a non-compact Calabi-Yau threefold. The topological vertex completely solved the problem of determining the usual topological string amplitudes for non-compact toric Calabi-Yau threefolds \cite{AKMV}. The refined topological vertex formulation extends the usual one and calculates the refined amplitudes for a certain class of non-compact toric Calabi-Yau threefolds \cite{IKV}.  The class of Calabi-Yau threefolds for which the refined vertex formalism works includes the geometries giving rise to gauge theory in spacetime via geometric engineering \cite{Katz:1996fh}. However, there exist toric Calabi-Yau threefolds which do not give rise to a gauge theory, such as the total space of the ${\cal O}(-3)$ bundle over $\mathbb{P}^2$ also known as local $\mathbb{P}^{2}$, and the refined topological vertex formalism can not be applied to such geometries directly\footnote{However, BPS multiplicities can be calculated by appropriately identifying the curves that survive the blow down from a larger geometry, to which refined vertex formalism can be applied, in which it can be embedded. For local $\mathbb{P}^2$ the BPS multiplicities were calculated in \cite{IKV}}. 

In this paper we define a new two legged vertex which can be used along with the refined vertex to calculate the partition function of not only the local $\mathbb{P}^2$ but a class of geometries which contain $\mathbb{P}^2$ as a divisor. We define this new vertex using the refined partition function of Hopf link \cite{GIKV} and flop transition. The partition function of local $\mathbb{P}^2$ was also obtained in \cite{AS2} using refined Chern-Simons theory \cite{AS}. 

The paper is organized as follows: In section 2, we discuss the refined topological vertex formalism and the appearance of a preferred direction. In section 3, we discuss the case of local $\mathbb{P}^2$ and the need for a new vertex. In section 3 we discuss properties of the refined partition function of Hopf link and show that it can be used to determine the refined partition function of local $\mathbb{P}^2$. In this section we also determine the a new two legged vertex which extends the refined vertex formalism. In section 4 we discuss our conclusions and future directions.

\section{The Refined Vertex Formalism}

The refinement of topological strings has been studied both from the A-model side as well as the B-model point of view
 \cite{HIV, IKV, AK1, DV, Hohenegger, KW, ACDKV, B4, AS, AS2, ONtalk}. On the A-model side the refined topological vertex provides a powerful method for calculating the open and the closed partition functions for a certain class of toric Calabi-Yau threefolds \cite{IKV}. The geometries for which the refined vertex formalism can be used are essentially those which give rise to ${\cal N}=2$ gauge theories in four dimensions via geometric engineering. Some simple Calabi-Yau threefolds such as local ${\mathbb P}^2$ (total space of ${\cal O}(-3)$ bundle on ${\mathbb P}^2$) do not give rise to gauge theory upon type IIA compactification. It is not possible to directly use the refined topological vertex formalism to write down the partition function for this Calabi-Yau threefold. In \cite{IKV} quantum numbers of the BPS states coming from branes wrapping holomorphic curves in  local $\mathbb{P}^2$ were determined by studying curves in a blowup of local $\mathbb{P}^{2}$ using the fact that refined topological vertex formalism can be applied to it.\footnote{Blowup of local $\mathbb{P}^2$ gives rise to $SU(2)$ gauge theory and refined vertex formalism can be used to write down its partition function \cite{IKV}. }.

In the refined vertex formalism the Calabi-Yau geometry is divided in ${\mathbb C}^{3}$ patches and the partition function is computed by suitably gluing the patches. This is exactly as in the usual topological vertex formalism. The difference arises in how the vertices are glued since the refined topological vertex, labelled by three Young diagrams, is a function of two parameters $(q,t)=(e^{i\epsilon_{1}},e^{-i\epsilon_{2}})$ \cite{IKV} (see Appendix A for notation and conventions being used):
\bea\nonumber
C_{\lambda\,\mu\,\nu}(t,q)&=&\Big(\frac{q}{t}\Big)^{\frac{\Arrowvert\mu\Arrowvert^2}{2}}\,t^{\frac{\kappa(\mu)}{2}}\,q^{\frac{\Arrowvert\nu\Arrowvert^2}{2}}\,\widetilde{Z}_{\nu}(t,q)
\sum_{\eta}\Big(\frac{q}{t}\Big)^{\frac{|\eta|+|\lambda|-|\mu|}{2}}\,s_{\lambda^{t}/\eta}(t^{-\rho}\,q^{-\nu})\,s_{\mu/\eta}(t^{-\nu^t}\,q^{-\rho}),
\eea
where $s_{\lambda/\eta}(\mathbf{x})$ is the skew-Schur function and $\widetilde{Z}_{\nu}(t,q)$ is given by
\bea\nn
\widetilde{Z}_{\nu}(t,q)&=&\prod_{(i,j)\in \nu}\Big(1-q^{a(i,j)}\,t^{\ell(i,j)+1}\Big)^{-1}\,.
\eea
The function $C_{\lambda\,\mu\,\nu}(t,q)\times M(t,q)$, where $M(t,q)=\prod_{i,j=1}^{\infty}(1-q^{i}\,t^{j-1})^{-1}$ is a refinement of the MacMahon function, is a combinatorial object just like the usual topological vertex \cite{ORV}. $C_{\lambda\,\mu\,\nu}(t,q)\times M(t,q)$ is the generating function of 3D partitions counting the partitions in an anisotropic way \cite{IKV}:
\bea
C_{\lambda\,\mu\,\nu}(t,q)\times M(t,q)=\sum_{\pi(\lambda\,\mu\,\nu)}q^{|\pi(\lambda\,\mu\,\nu)|_{\nu}}\,t^{|\pi(\lambda\,\mu\,\nu)|-|\pi(\lambda\,\mu\,\nu)|_{\nu}},
\eea
where the sum is over 3D partitions $\pi(\lambda\,\mu\,\nu)$ with asymptotics determined by $\lambda,\mu$ and $\nu$ along the three directions and $|\pi(\lambda\,\mu\,\nu)|$ is the regularized total number of boxes in $\pi(\lambda\,\mu\,\nu)$ \cite{ORV}. The shape of the partition $\nu$ along the z-axis determines a slicing of the 3D partition by planes parallel to the z-axis. For the usual vertex this slicing of the 3D partition is just a convenient way of being able to calculate the generating function and the result at the end is independent of this slicing. For the refined vertex case, however, this is not the case and the direction along which we slice breaks the cyclic symmetry of the vertex. This direction along which we slice we will call {\it the preferred direction}. Some slices are labelled by $q$ and some slices are labelled by $t$ depending on the shape of the partition $\nu$.  $|\pi(\lambda\,\mu\,\nu)|_{\nu}$ is the number of boxes that fall on the $q$-slices. Each slice is assigned one of the two counting parameters $q$ or $t$ depending on whether this particular slice is located between an inner,$\{v_{i}\}$, and outer corner ,$\{u_{i}\} $, or between an outer and an inner while we trace the profile of $\nu$, \figref{slicing}.
\begin{figure}[h]
  \centering
  \includegraphics[width=2.5in]{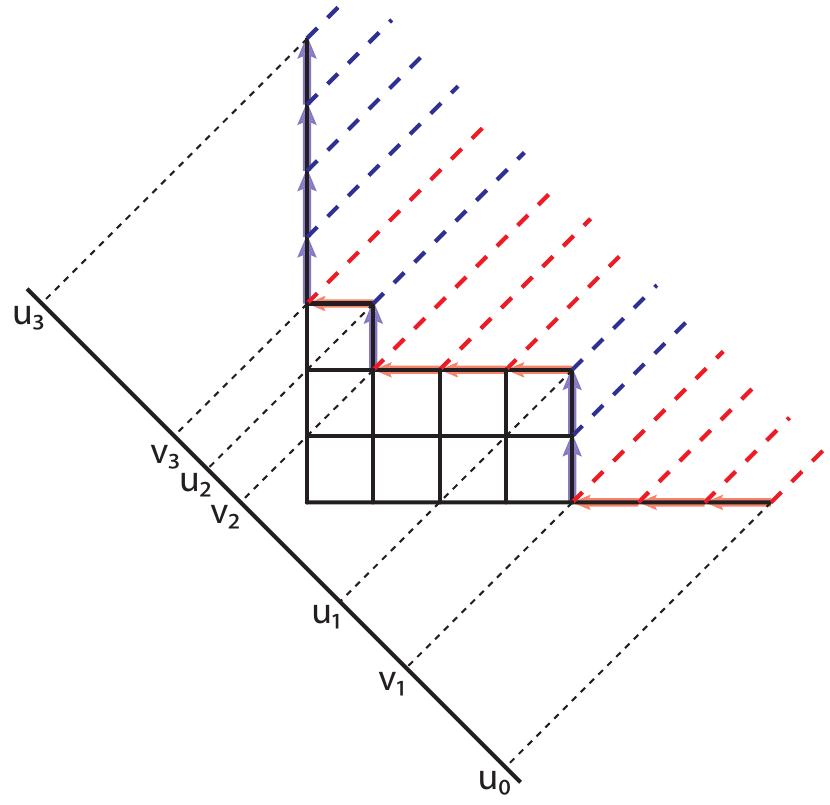}\\
  \caption{The counting scheme is uniquely determined by the representation along the preferred direction.}\label{slicing}
\end{figure}

\noindent If all representations labeling the vertex are trivial, the counting is depicted in \figref{macmahon}, and the refined MacMahon function is the generating function of 3D partition with respect to this counting.

\begin{figure}[h]
  \centering
  \includegraphics[width=2.4in]{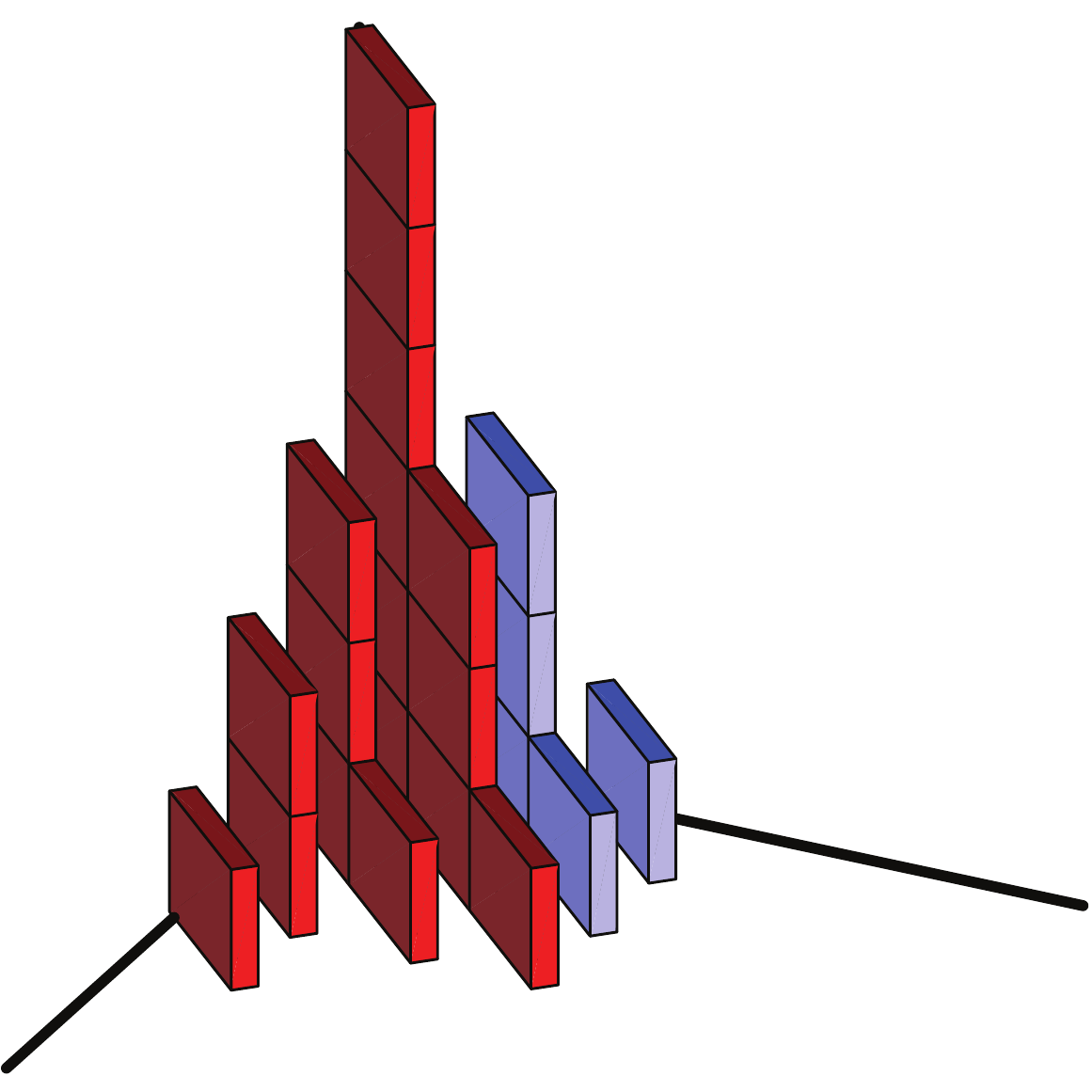}\\
  \caption{Red slices (left) are counted with $t$ and the blue slices (right) with $q$.}\label{macmahon}
\end{figure}

For any partition $\nu$ (along the z-axis), it is clear that slices in the $x$ and $y$ direction (far enough) are counted with either $q$ alone or  counted with $t$ alone.  Thus we see that the two un-preferred directions are either associated with $t$ or with $q$, \textit{i.e.}, with $\varepsilon_{1}$ or $\varepsilon_{2}$.


\begin{figure}[h]
  \centering
  \includegraphics[width=3in]{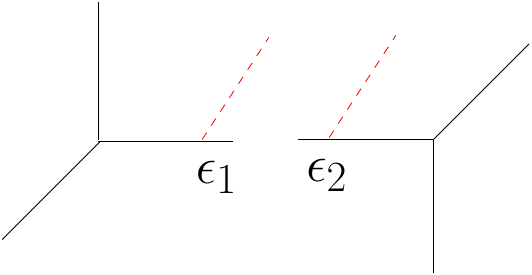}\\
  \caption{Gluing of two refined vertices.}\label{glue}
\end{figure}

When gluing two refined vertices together along an un-preferred direction we must have the edge of one vertex associated with $\varepsilon_{1}$ and the edge of the second vertex associated with $\varepsilon_{2}$, as shown in \figref{glue}. If this is not the case the symmetry between $q$ and $t$ in the refined partition function, which should hold on general grounds \cite{HIV}, is lost.

Also since at each vertex we have a preferred edge, along which we slice in order to define the refined vertex, in the web diagram we will have a number of preferred edges (one for each vertex). Unless all these preferred edges are parallel to each other, i.e., we are slicing all the 3D partitions that appear at the vertices in the same way, the different vertices can not be glued correctly. Hence, a very important constraint of the refined vertex formalism is that all preferred edges (which define the preferred directions for the vertices connected by the edge) in the web diagram should be parallel. As we will discuss in the next section this condition can not always be satisfied and, therefore, refined vertex formalism only works for a certain class of geometries. An example of a geometry for which the corresponding web has a set of parallel preferred directions for each vertex is resolved $A_n$ singularity fibered over a $\mathbb{P}^1$ or a chain of $\mathbb{P}^1$'s.

\section{The Case of Local $\mathbb{P}^2$}

In order to calculate the refined partition function using the refined vertex we need to choose for every vertex of the web diagram an edge which is associated with the direction we use to slice the 3D partition. This choice is arbitrary for a given vertex, however, all such edges in the web diagram need to be parallel to each other. This essentially means that we need a set of edges of the web diagram such that this set covers all the vertices and all  edges in this set are parallel. It is easy to see that many web diagram fail to have such a set of edges and therefore refined vertex formalism can not be directly applied to such web diagrams.

The simplest geometry which can not be dealt with using the refined topological vertex is the total space of the line bundle ${\cal O}(-3) \mapsto \mathbb{P}^{2}$ also referred to as local $\mathbb{P}^{2}$. The Newton polygon and the web diagram of this geometry is given in \figref{p2} below.

\begin{figure}[h]
  \centering
  \includegraphics[width=4in]{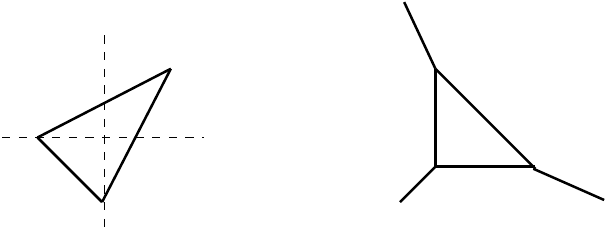}\\
  \caption{The Newton polygon and the web diagram of local $\mathbb{P}^{2}$.}\label{p2}
\end{figure}

If we choose the horizontal internal edge of of the web diagram (\figref{p2}) of local $\mathbb{P}^{2}$ to be the preferred edge (as shown in \figref{p22}) then the refined vertex factor associated with this edge is given by,
\bea\label{eq1}
C_{\lambda^t\,\emptyset\,\nu}(t,q)\,C_{\emptyset\,\mu\,\nu^t}(q,t)\sim \widetilde{Z}_{\nu}(t,q)\widetilde{Z}_{\nu^t}(q,t)\,\Big[s_{\lambda}(t^{-\rho}q^{-\nu})\,s_{\mu}(t^{-\rho}q^{-\nu})\Big]\,.
\eea
\begin{figure}[h]
  \centering
  \includegraphics[width=1.2in]{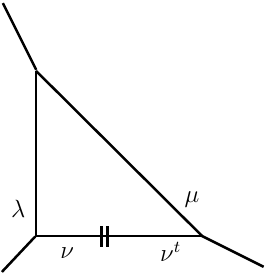}\\
  \caption{The labeling of the edges with horizontal preferred direction.}\label{p22}
\end{figure}

For $\nu$ a trivial partition we see that the factor in the square bracket in Eq.(\ref{eq1}) is such that both partitions $\lambda$ and $\mu$ appear with parameter $t$ (i.e.,\,the parameter $\varepsilon_{2}$). Therefore this suggests the assignment of parameters $\varepsilon_1$ and $\varepsilon_2$ for the local $\mathbb{P}^2$ web diagram as shown in \figref{localp2slicing}.

\begin{figure}[h]
  \centering
  \includegraphics[width=4.5in]{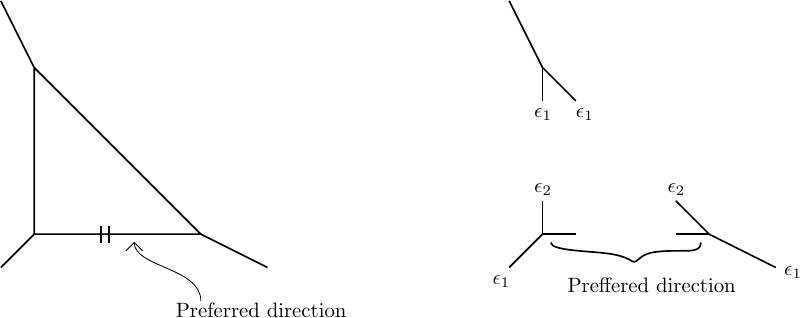}\\
  \caption{Assignment of $\epsilon_{1,2}$ parameters to the edges in the local $\mathbb{P}^2$ web diagram.}\label{localp2slicing}
\end{figure}

Thus we see that the upper vertex has a different brane configuration then the lower ones which had different parameters on the un-preferred edges. Thus we need a new vertex in which two of the edges are associated with the same parameter. Thus to apply the refined vertex formalism to local $\mathbb{P}^2$ we need two refined topological vertices corresponding to two assignment of parameters to the un-preferred directions as shown in \figref{twovertex}.

\begin{figure}[h]
  \centering
  \includegraphics[width=3in]{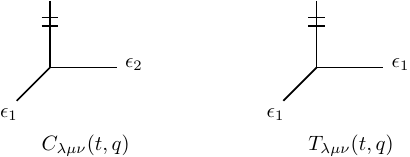}\\
  \caption{The two vertices needed for local $\mathbb{P}^2$.}\label{twovertex}
\end{figure}

\section{The Refined Hopf Link and the New Vertex}

In this section, we will calculate the refined partition function of local $\mathbb{P}^2$ and, along the way, determine the two-legged version of the new refined vertex.

We will begin with local $\mathbb{P}^{2}$ blown up at one point. The blowup of $\mathbb{P}^{2}$ is a Hirzebruch surface \textit{i.e.}, it is a $\mathbb{P}^{1}$ bundle over $\mathbb{P}^{1}$ which is the projectivization of the bundle ${\cal O}_{\mathbb{P}^1}(-1)\oplus {\cal O}_{\mathbb{P}^1}(0)$ and will be denoted by $\mathbb{F}_{1}$. The local $\mathbb{F}_{1}$ is then the total space of the canonical bundle on $\mathbb{F}_{1}$. The Newton polygon and the web diagram of local $\mathbb{F}_{1}$ is shown in \figref{hirzebruch1}.

\begin{figure}[h]
  \centering
  \includegraphics[width=3in]{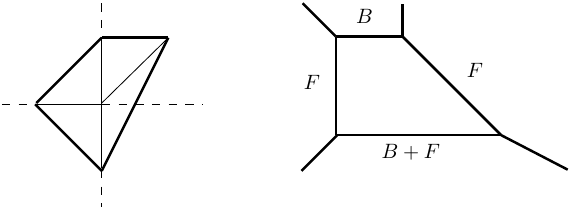}\\
  \caption{The Newton polygon and the web diagram of local $\mathbb{F}_{1}$.}\label{hirzebruch1}
\end{figure}

We denote by $B$ and $F$ the homology class of the base and the fiber $\mathbb{P}^{1}$ respectively, these have the following intersection numbers,
\bea
B\cdot B=-1\,,\,\,\,F\cdot F=0\,,\,\,\,\,B\cdot F=+1\,.
\eea
For a $\mathbb{P}^{1}$ embedded in a Calabi-Yau threefold the local geometry is ${\cal O}_{\mathbb{P}^{1}}(m)\oplus {\cal O}_{\mathbb{P}^{1}}(-2-m)$ for some $m$. In the case of the local $\mathbb{F}_{1}$ $B$ is the exceptional curve coming from the blowup of local $\mathbb{P}^{2}$ and has local geometry ${\cal O}_{\mathbb{P}^1}(-1)\oplus {\cal O}_{\mathbb{P}^{1}}(-1)$ i.e., it is locally resolved conifold. The curve $B$ can undergo flop transition which is also clear from the Newton polygon in \figref{hirzebruch1} which has two distinct triangulations. The flop transition of $B$ can take us from local $\mathbb{F}_{1}$ to local $\mathbb{P}^{2}$ as shown in \figref{flop}.

\begin{figure}[h]
\begin{center}
\includegraphics[width=3in]{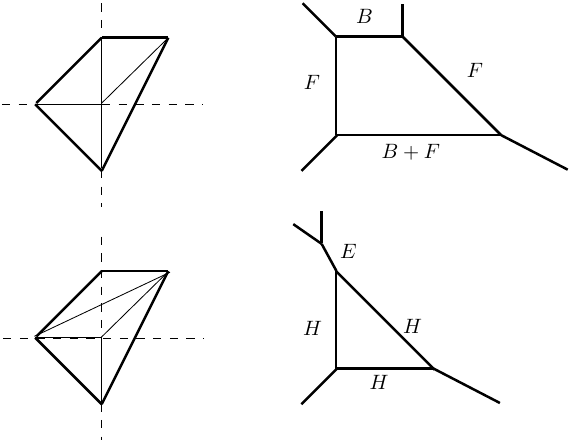}
\end{center}
\caption{The flop of local $\mathbb{F}_{1}$ gives rise to local $\mathbb{P}^{2}$ when the flopped curve is taken to infinite size.}\label{flop}
\end{figure}

The curve $E$ comes from the flop of $B$ and $H$ is the hyperplane class of $\mathbb{P}^{2}$ and is given by $B+F$. The relation between K\"ahler parameters on the two sides of the flop is given by
\bea
Q_{E}=Q_{B}^{-1}\,,\,\,\,Q_{H}=Q_{B}Q_{F}\,,
\eea
where $Q_{E}=e^{-t_{E}}$ is the parameter associated with the curve $E$.

From \figref{flop} it is clear that we can obtain the refined partition function of local $\mathbb{P}^{2}$ from the refined partition function of local $\mathbb{F}_{1}$, $Z_{\mathbb{F}_{1}}(Q_{B},Q_{F})$, if we can follow flop transition in the partition functions,
\bea\label{limit}
Z_{\mathbb{P}^{2}}(Q_{H})=\lim_{Q_{E}\mapsto 0}\, Z_{\mathbb{F}_{1}}(Q_{E}^{-1},Q_{H}Q_{E})\,.
\eea
The refined partition function of the local $\mathbb{F}_{1}$ can be calculated using the refined topological vertex, the two internal horizontal line in the web diagram of the local $\mathbb{F}_{1}$ (\figref{hirzebruch1}) can serve as the preferred edges, and is given by \cite{IKV},
\begin{align}\nn
Z_{\mathbb{F}_{1}}(Q_{B},Q_{F}):&=\sum_{\nu\,\lambda\,\mu\,\sigma}(-Q_{B}Q_{F})^{|\nu|}(-Q_B)^{|\sigma|}(-Q_{F})^{|\lambda|+|\mu|}\,f_{\nu}^{2}(q,t)\,\widetilde{f}_{\lambda}(t,q)\,\widetilde{f}_{\mu}(q,t)\\
&\times C_{\emptyset\, \lambda\,\nu^{t}}(t,q)\,C_{\mu^{t}\,\emptyset\,\nu}(q,t)\,C_{\lambda^{t}\,\emptyset\,\sigma}(t,q)\,C_{\emptyset\,\mu\,\sigma^{t}}(q,t),
\end{align}
where we are using the following convention for the framing factors\footnote{This convention eliminates the need for shifting certain K\"ahler parameters by $\frac{\varepsilon_{1}+\varepsilon_{2}}{2}$.}
\begin{align}\nonumber
f_{\lambda}(t,q)=(-1)^{|\lambda|}\,t^{\frac{\Arrowvert\lambda^t\Arrowvert^2}{2}}\,q^{-\frac{\Arrowvert\lambda\Arrowvert^2}{2}}\,,\,\,\,\,\,\widetilde{f}_{\lambda}(t,q)=(-1)^{|\lambda|}\,t^{\frac{\Arrowvert\lambda^t\Arrowvert^2-|\lambda|}{2}}\,q^{-\frac{\Arrowvert\lambda\Arrowvert^2-|\lambda|}{2}}.
\end{align}
After the flop transition the relevant variables are $Q_{E}=Q_{B}^{-1}$ and $Q_{H}=Q_{B}Q_{F}$. In terms of these variables the partition function is,
\begin{align}\nonumber
Z_{\mathbb{F}_{1}}(Q_{E}^{-1},Q_{H}Q_{E})&=\sum_{\nu\,\lambda\,\mu\,\sigma}(-Q_{H})^{|\lambda|+|\mu|+|\nu|}(-Q_{E})^{-|\sigma|}\,Q_{E}^{|\lambda|+|\mu|}\,f_{\nu}^{2}(q,t)\,\widetilde{f}_{\lambda}(t,q)\,\widetilde{f}_{\mu}(q,t)\\
&\times C_{\emptyset\, \lambda\,\nu^{t}}(t,q)\,C_{\mu^{t}\,\emptyset\,\nu}(q,t)\,C_{\lambda^{t}\,\emptyset\,\sigma}(t,q)\,C_{\emptyset\,\mu\,\sigma^{t}}(q,t).
\end{align}

Separating the sum involving $Q_{E}$, we can write the partition function as
\begin{align}\nonumber
Z_{\mathbb{F}_{1}}(Q_{E}^{-1},Q_{H}Q_E)&=\sum_{\nu\,\lambda\,\mu\,\sigma}(-Q_{H})^{|\lambda|+|\mu|+|\nu|}\,f_{\nu}^{2}(q,t)\,\widetilde{f}_{\lambda}(t,q)\,\widetilde{f}_{\mu}(q,t)\,C_{\emptyset\, \lambda\,\nu^{t}}(t,q)\,C_{\mu^{t}\,\emptyset\,\nu}(q,t)\\\label{f1}
&\times\,\, Q_{E}^{|\lambda|+|\mu|}\,Z_{\lambda\,\mu}(Q_{E}^{-1}),
\end{align}
where,
\begin{align}\label{rhl}
Z_{\lambda\,\mu}(Q)=\sum_{\sigma}(-Q)^{|\sigma|}\,C_{\lambda^{t}\,\emptyset\,\sigma}(t,q)\,C_{\emptyset\,\mu\,\sigma^{t}}(q,t).
\end{align}
$Z_{\lambda\,\mu}(Q)$ is the the refined Hopf link invariant studied in \cite{GIKV}. It is the refinement of the Hopf link invariant in which the two linked unknots are colored by $\lambda$ and $\mu$ (\figref{hopf3}). 

\begin{figure}[h]
\begin{center}
\includegraphics[width=5in]{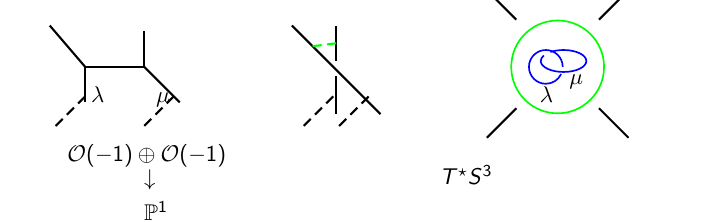}
\end{center}
\caption{The geometric transition between the resolved conifold with Lagrangian branes and $T^{*}S^3$ with colored Hopf link.}\label{hopf3}
\end{figure}

The unrefined Hopf link invariant can be calculated using the Chern-Simons theory \cite{OV} and the refined Hopf link can be calculated using either the refined vertex \cite{GIKV} or the refined Chern-Simons theory \cite{AS}. $Z_{\lambda\,\mu}(Q^{-1})$ and the limit given in Eq.(\ref{limit}) can be determined using certain properties of the Macdonald polynomials. 

\subsection{Flop Transition and Refined Hopf Link}
Although $Z_{\lambda\,\mu}(Q,t,q)$ is an infinite series in $Q$, the normalized open string partition function
\bea
G_{\lambda\,\mu}(Q)=\frac{Z_{\lambda\,\mu}(Q,t,q)}{Z_{\emptyset\,\emptyset}(Q,t,q)}=\frac{\sum_{\sigma}(-Q)^{|\sigma|}\,C_{\lambda^{t}\,\emptyset\,\sigma}(t,q)\,C_{\emptyset\,\mu\,\sigma^{t}}(q,t)}{\sum_{\sigma}(-Q)^{|\sigma|}\,C_{\emptyset\,\emptyset\,\sigma}(t,q)\,C_{\emptyset\,\emptyset\,\sigma^{t}}(q,t)}
\eea
is a polynomial of degree $|\lambda|+|\mu|$ in $Q$. This was conjectured in \cite{GIKV} and proven in \cite{AK}. Notice that $Z_{\emptyset\,\emptyset}(Q)$ is the closed string partition function of the geometry ${\cal O}_{\mathbb{P}^{1}}(-1)\oplus{\cal O}_{\mathbb{P}^{1}}(-1)$ and is invariant under the flop,\footnote{It is invariant if we ignore the shift in the classical contribution which arises due to the flop.}
\bea
Z_{\emptyset\,\emptyset}(Q^{-1},t,q)=Z_{\emptyset\,\emptyset}(Q,t,q)\,.
\eea
Thus we can obtain the expression for the flopped partition function by studying $G_{\lambda\,\mu}(Q,t,q)$. Using the definition of the refined topological vertex we can write $G_{\lambda\,\mu}(Q,t,q)$ as
\bea\nonumber
G_{\lambda\,\mu}(Q)&=&\frac{q^{\frac{|\lambda|}{2}}t^{-\frac{|\lambda|}{2}}\widetilde{f}_{\mu^t}\sum_{\nu}\,(-Q)^{|\nu|}\,q^{\frac{\Arrowvert\nu\Arrowvert^2}{2}}\,
t^{\frac{\Arrowvert\nu^t\Arrowvert^2}{2}}\,\widetilde{Z}_{\nu}(t,q)\,\widetilde{Z}_{\nu^t}(q,t)\,s_{\lambda}(t^{-\rho}q^{-\nu})\,
s_{\mu}(t^{-\rho}q^{-\nu})}{(-1)^{|\mu|}\sum_{\nu}\,(-Q)^{|\nu|}\,q^{\frac{\Arrowvert\nu\Arrowvert^2}{2}}\,t^{\frac{\Arrowvert\nu^t\Arrowvert^2}{2}}\,\widetilde{Z}_{\nu}(t,q)\,\widetilde{Z}_{\nu^t}(q,t)}\,.
\eea
Notice that the two Schur functions in the summation above have the same argument therefore we can express their product in terms of the Macdonald polynomials,
\bea\label{basischange}
s_{\lambda}({\bf x})\,s_{\mu}({\bf x})=\sum_{\eta}N^{\eta}_{\lambda\,\mu}s_{\eta}({\bf x})=\sum_{\eta\,\sigma}N^{\eta}_{\lambda\,\mu}U_{\eta\sigma}P_{\sigma}({\bf x}),
\eea
where the $U_{\eta\,\sigma}(q,t)$ are the previously defined matrices which take Macdonald functions to Schur functions.
$s_{\eta}({\bf x})$ and $P_{\sigma}({\bf x};q,t)$ are both homogenous functions of degree of $|\eta|$ and $|\sigma|$, respectively, hence, $U_{\eta\,\sigma}(q,t)=0$ if $|\eta|\neq |\sigma|$. Also it follows from the definition of the Macdonald function that $U_{\eta\,\sigma}=\delta_{\eta\,\sigma}$ for $t=q$.

Using Eq.(\ref{basischange}) we can express $G_{\lambda\,\mu}$ as
\bea
G_{\lambda\,\mu}(Q,t,q)=(-1)^{|\mu|}q^{\frac{|\lambda|}{2}}\,t^{-\frac{|\lambda|}{2}}\widetilde{f}_{\mu^t}\sum_{\eta\,\sigma}\,N^{\eta}_{\lambda\,\mu}
U_{\eta\sigma}\Big(\frac{J_{\sigma}(Q,t,q)}{J_{\emptyset}(Q,t,q)}\Big)\,,
\eea
where,
\bea
J_{\sigma}(Q,t,q)=\sum_{\nu}\,(-Q)^{|\nu|}\,q^{\frac{\Arrowvert\nu\Arrowvert^2}{2}}\,t^{\frac{\Arrowvert\nu^t\Arrowvert^2}{2}}\,\widetilde{Z}_{\nu}(t,q)\,\widetilde{Z}_{\nu^t}(q,t)\,
P_{\sigma}(t^{-\rho}q^{-\nu};q,t),
\eea
The principal specialization of the Macdonald polynomials are proportional to $\widetilde{Z}_{\nu}(t,q)$,
\bea
t^{\frac{\Arrowvert\nu^t\Arrowvert^2}{2}}\,\widetilde{Z}_{\nu}(t,q)&=&P_{\nu}(t^{-\rho};q,t),
\eea
which allows us to express $J_{\sigma}(Q,t,q)$ in terms of Macdonald functions,
\bea
J_{\sigma}(Q,t,q)&=&\sum_{\nu}\,(-Q)^{|\nu|}\,\,P_{\nu^t}(q^{-\rho};t,q)\,P_{\nu}(t^{-\rho};q,t)\,
P_{\sigma}(t^{-\rho}q^{-\nu};q,t).
\eea
The sum over the partition $\nu$ can be performed by making use of the identity (\cite{macdonald}, Page 332, Eq.(6.6))
\bea
P_{\nu}(t^{-\rho};q,t)P_{\sigma}(q^{-\nu}\,t^{-\rho};q,t)=P_{\sigma}(t^{-\rho};q,t)\,P_{\nu}(q^{-\sigma}\,t^{-\rho};q,t).
\eea
Using the above identity $J_{\sigma}(Q,t,q)$ takes a product form,
\bea\nonumber
J_{\sigma}(Q,t,q)&=&\sum_{\nu}\,(-Q)^{|\nu|}\,P_{\nu^t}(q^{-\rho};t,q)\,P_{\sigma}(t^{-\rho};q,t)\,P_{\nu}(t^{-\rho}q^{-\sigma};q,t)\\\nonumber
&=&P_{\sigma}(t^{-\rho};q,t)\sum_{\nu}\,(-Q)^{|\nu|}\,P_{\nu^t}(q^{-\rho};t,q)\,P_{\nu}(t^{-\rho}q^{-\sigma};q,t)\\
&=&P_{\sigma}(t^{-\rho};q,t)\,\prod_{i,j=1}^{\infty}\left(1-Q\,t^{-\rho_{i}}\,q^{-\rho_{j}-\sigma_{i}}\right),
\eea
where in the last line above we have used the identity \cite{macdonald},
\bea
\sum_{\nu}\,P_{\nu}({\bf x};q,t)\,P_{\nu^t}({\bf y};t,q)=\prod_{i,j=1}^{\infty}\left(1+x_{i}\,y_{j}\right).
\eea
We can factorize the infinite product as a finite product over the boxes of a Young diagram and an infinite one to normalize $J_{\sigma}(Q,t,q)$:
\begin{align}
&J_{\sigma}(Q,t,q)=\left[P_{\sigma}(t^{-\rho};q,t)\,\prod_{(i,j)\in \sigma}(1-Q\,t^{i-\frac{1}{2}}\,q^{-j+\frac{1}{2}})\right]\times \prod_{i,j=1}^{\infty}\left(1-Q\,t^{i-\frac{1}{2}}\,q^{j-\frac{1}{2}}\right)\\\nn
&\frac{J_{\sigma}(Q,t,q)}{J_{\emptyset}(Q,t,q)}=P_{\sigma}(t^{-\rho};q,t)\,\prod_{(i,j)\in \sigma}(1-Q\,t^{i-\frac{1}{2}}\,q^{-j+\frac{1}{2}}).
\end{align}
Thus the normalized refined Hopf link partition function can be written as,
\bea\label{xxxx}\nonumber
G_{\lambda\,\mu}(Q)=g_{\lambda\,\mu}\sum_{\eta\,\,\sigma}N^{\eta}_{\lambda\,\mu}U_{\eta\,\sigma}(q,t)\,P_{\sigma}(t^{-\rho};q,t)\,\prod_{(i,j)\in \sigma}(1-Q\,t^{i-\frac{1}{2}}\,q^{-j+\frac{1}{2}}),
\eea
where $g_{\lambda\,\mu}=(-1)^{|\mu|}\,q^{\frac{|\lambda|}{2}}t^{-\frac{|\lambda|}{2}}\widetilde{f}_{\mu^t}(t,q)$. Before we continue with the derivation of the new vertex, let us argue that the normalized partition function for the Hopf link is a polynomial in $Q$ of order $|\lambda|+|\mu|$. We know that $N^{\eta}_{\lambda\,\mu}$ is zero unless $|\lambda|+|\mu|=|\eta|$ and $U_{\eta\,\sigma}(t,q)$ vanishes for $|\eta|\neq|\sigma|$. Therefore, we conclude that the sum is finite and $G_{\lambda\,\mu}(Q,t,q)$ is of the order $|\lambda|+|\mu|$.

The two legged new vertex will be obtained by taking the flop transition and sending the K\"{a}hler parameter to infinity. Let us define $Q_{\bullet}=Q^{-1}$ so that, 
\bea
\prod_{(i,j)\in \sigma}\Big(1-Q\,t^{i-\frac{1}{2}}\,q^{-j+\frac{1}{2}}\Big)&=&\prod_{(i,j)\in \sigma}\Big(1-Q_{\bullet}^{-1}\,t^{i-\frac{1}{2}}\,q^{-j+\frac{1}{2}}\Big)\,,\\\nonumber
&=&(-Q_{\bullet})^{-|\sigma|}\,t^{\frac{\Arrowvert\sigma^t\Arrowvert^2}{2}}\,q^{-\frac{\Arrowvert\sigma\Arrowvert^2}{2}}\,\prod_{(i,j)\in \sigma}\left(1-Q_{\bullet}\,t^{-i+\frac{1}{2}}\,q^{j-\frac{1}{2}}\right)\,.
\eea
The limit to define the two-legged vertex (up to framing and overall factors) can be easily taken using the following form of the normalized Hopf link partition function,
\begin{align}\label{xx}\nonumber
G_{\lambda\,\mu}(Q)
&=(-1)^{|\lambda|}\,Q^{|\lambda|+|\mu|}q^{\frac{|\lambda|}{2}}t^{-\frac{|\lambda|}{2}}\widetilde{f}_{\mu^t}(t,q)\sum_{\eta\,\,\sigma}N^{\eta}_{\lambda\,\mu}U_{\eta\,\sigma}(q,t)\,t^{\frac{||\sigma^t||^2}{2}}\,q^{-\frac{||\sigma||^2}{2}}\,P_{\sigma}(t^{-\rho};q,t)\\ &\times\prod_{(i,j)\in \sigma}\left(1-Q^{-1}\,t^{-i+\frac{1}{2}}\,q^{j-\frac{1}{2}}\right).
\end{align}
From the above equation we see that
\bea
\lim_{Q\mapsto 0}\,Q^{|\lambda|+|\mu|}\,Z_{\lambda\,\mu}(Q^{-1})&=&\lim_{Q\mapsto 0}\,Q^{|\lambda|+|\mu|}\,Z_{\emptyset\,\emptyset}(Q^{-1})\,G_{\lambda\,\mu}(Q^{-1})\\\nonumber
&=&(-1)^{|\mu|}q^{\frac{|\lambda|}{2}}t^{-\frac{|\lambda|}{2}}\widetilde{f}_{\mu^t}(t,q)\sum_{\eta\,\,\sigma}N^{\eta}_{\lambda\,\mu}U_{\eta\,\sigma}(q,t)\,f_{\sigma}(t,q)\,P_{\sigma}(t^{-\rho};q,t).
\eea
Notice that we have made use of the invariance of the resolved conifold partition function under flop, $Z_{\emptyset\,\emptyset}(Q^{-1},t,q)=Z_{\emptyset\,\emptyset}(Q,t,q)$. Using the above in Eq.(\ref{limit}) and Eq.(\ref{f1}) we get the following expression for the refined partition function of local $\mathbb{P}^{2}$:\\
\vskip 0.05cm
\fbox{
\begin{minipage}{5.5in}
\bea\label{RPF}
&&Z_{\mathbb{P}^2}(Q_{H})=\sum_{\lambda\,\mu\,\nu}(-Q_{H})^{|\lambda|+|\mu|+|\nu|}\,q^{\frac{3||\nu^t||^2}{2}}\,t^{-\frac{||\nu||^2}{2}}\,\widetilde{Z}_{\nu}(q,t)
 \widetilde{Z}_{\nu^t}(t,q)\times\\\nonumber&&s_{\lambda}(q^{-\rho}t^{-\nu})\,s_{\mu}(q^{-\rho}t^{-\nu})\,\Big(\frac{q}{t}\Big)^{\frac{|\lambda|-|\mu|}{2}}\,
 \Big[\sum_{\eta\,\sigma}\,N^{\eta}_{\lambda\,\mu}\,U_{\eta\,\sigma}t^{\frac{||\sigma^t||^2}{2}}q^{-\frac{||\sigma||^2}{2}}\,P_{\sigma}(t^{-\rho};q,t)\Big]\,.
\eea
\end{minipage}}
\vskip 0.3cm
To identify the new vertex we express Eq.(\ref{RPF}) as,
\bea
Z_{\mathbb{P}^2}(Q_{H})&=&\sum_{\lambda\,\mu\,\nu}(-Q_{H})^{|\lambda|+|\mu|+|\nu|}\,
\Big(f_{\nu}(q,t)\,\widetilde{f}_{\lambda}(t,q)\,\widetilde{f}_{\mu}(q,t)\Big)^{2}\\\nn
&&\,C_{\emptyset\,\lambda\,\nu^t}(t,q)C_{\mu^t\,\emptyset\,\nu}(q,t)\,T_{\mu\,\lambda^t\,\emptyset}(t,q)\,,
\eea
with $T_{\mu\lambda\emptyset}$ being the new two legged vertex. Comparing with Eq.(\ref{RPF}) gives the following expression for $T_{\lambda\mu\emptyset}$,\\
\vskip 0.05cm
\fbox{
\begin{minipage}{5.5in}
\bea
T_{\lambda\,\mu\,\emptyset}(t,q)=(-1)^{|\mu|}\,f_{\mu}(q,t)\,f^{2}_{\lambda^t}(t,q)\,
\sum_{\eta\,\,\sigma}N^{\eta}_{\lambda\,\mu^t}U_{\eta\,\sigma}(q,t)\,
f_{\sigma}\,P_{\sigma}(t^{-\rho};q,t)\,.
\eea
\end{minipage}}
\vskip 0.3cm
We have verified, up to order $Q_{H}^5$, that the partition function in Eq.(\ref{RPF}) gives the same BPS degeneracies as were obtained in \cite{IKV} from studying curves in local $\mathbb{F}_{1}$ which blow down to local $\mathbb{P}^2$.

\section{Conclusions}
We have defined a new two legged vertex which can be used along with the refined vertex to calculate the refined partition function of local $\mathbb{P}^2$. Beyond just local $\mathbb{P}^2$ this new vertex, together with the refined vertex, allows us the calculate the refined partition function for Calabi-Yau geometries which are obtained from the resolution of the orbifold $\mathbb{C}^3/\mathbb{Z}_{2k+1}$. All these geometries contain $\mathbb{P}^2$ as a divisor. In this paper we have only obtained the two legged vertex, however, a vertex with all three partitions non-trivial can also be obtained as was done in \cite{toappear}. But there are ambiguities in identifying the three legged vertex which leads to a hierarchy of vertices \cite{toappear} which might be related to the ``index vertex" of Okounkov and Nekrasov \cite{ONtalk}. The new vertex we defined in this paper is also related to refined Chern-Simons theory by a change of basis of holonomies \cite{toappear}.

\noindent\section*{ Acknowledgments}
We would like to thank Mina Aganagic, Johannes Walcher and Cumrun Vafa for many useful discussions. We would also like to thank the Simon Center for Geometry and Physics for hospitality where part of this work was carried out. AI was partly supported by the  Higher Education Commission grant HEC-20-2518.

\section{Appendix A}
In this section, we want to review our notations, conventions and collect some of the useful formulas for the derivations. We want to warn the reader that the convention we follow in this note is different from the one used in \cite{IKV}. The difference comes from the pictorial representation of the Young diagrams. In  \cite{IKV}, for a given representation $\lambda=\{\lambda_{1},\lambda_{2},\mathellipsis \}$, we illustrated the corresponding Young diagram with columns of height $\lambda_{i}$. The arm length $a(i,j)$ and leg length $\ell(i,j)$ of a given box $(i,j)$ in the diagram were defined as the number of boxes to the right of that box and above that box, respectively. In this note we follow the more conventional notation shown in \figref{convention} where $\lambda_{i}$'s are the number of boxes in the $i^{th}$ row. The arm and leg lengths are defined accordingly.

\begin{figure}[h]
\begin{center}
\includegraphics[scale=0.7]{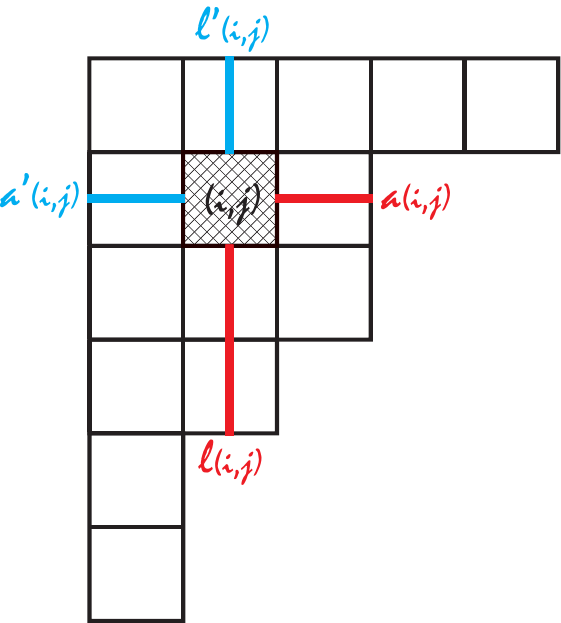}
\end{center}
\caption{The figure shows the present convention. $\lambda=\{5,3,3,2,1,1\}$}\label{convention}
\end{figure}
In addition to the arm and leg lengths, we define the co-arm length $a'(i,j)$ and co-leg length $\ell'(i,j)$
\begin{align}
&a(i,j)=\nu_{i}-j\,,\,\,\,\,\ell(i,j)=\nu^{t}_{j}-i\\
&a'(i,j)=j-1\,,\,\,\,\ell'(i,j)=i-1
\end{align}
Note that
\bea
a_{\lambda}(i,j)=\ell_{\lambda^{t}}(j,i),\qquad\qquad\ell_{\lambda}(i,j)=a_{\lambda^{t}}(j,i)
\eea
with $(i,j)\in\lambda$ and $(j,i)\in\lambda^{t}$.
The difference in the conventions reflects also on some identities used in \cite{IKV}, here we use
\begin{align}
&n(\lambda)=\sum_{i=1}^{\ell(\lambda)}(i-1)\lambda_{i}=\frac{1}{2}\sum_{i=1}^{\ell(\lambda)}\lambda^{t}_{i}(\lambda^{t}_{i}-1)=\sum_{(i,j)\in\lambda}\ell(i,j)=\sum_{(i,j)\in\lambda}\ell'(i,j)=\frac{\Arrowvert\lambda^{t}\Arrowvert^{2}}{2}-\frac{\lambda}{2}\\
&n(\lambda^{t})=\sum_{i=1}^{\ell(\lambda^{t})}(i-1)\lambda^{t}_{i}=\frac{1}{2}\sum_{i=1}^{\ell(\lambda^{t})}\lambda_{i}(\lambda_{i}-1)=\sum_{(i,j)\in\lambda}a(i,j)=\sum_{(i,j)\in\lambda}a'(i,j)=\frac{\Arrowvert\lambda\Arrowvert^{2}}{2}-\frac{\lambda}{2}
\end{align}
with $\ell(\lambda)$ being the number of non-zero $\lambda_{i}$'s, or in other words, it is the number of rows in the Young diagram. The hook length $h(i,j)$ and the content $c(i,j)$ are defined as
\bea
h(i,j)=a(i,j)+\ell(i,j)+1,\qquad c(i,j)=j-i
\eea
which satisfy
\begin{align}
&\sum_{(i,j)\in\lambda}h(i,j)=n(\lambda^{t})+n(\lambda)+|\lambda|,\\
&\sum_{(i,j)\in\lambda}c(i,j)=n(\lambda^{t})-n(\lambda).
\end{align}
The refined topological vertex in this convention is given by
\bea\nonumber
C_{\lambda\,\mu\,\nu}(t,q)&=&\Big(\frac{q}{t}\Big)^{\frac{||\mu||^2}{2}}\,t^{\frac{\kappa(\mu)}{2}}\,q^{\frac{||\nu||^2}{2}}\,\widetilde{Z}_{\nu}(t,q)
\sum_{\eta}\Big(\frac{q}{t}\Big)^{\frac{|\eta|+|\lambda|-|\mu|}{2}}\,s_{\lambda^{t}/\eta}(t^{-\rho}\,q^{-\nu})\,s_{\mu/\eta}(t^{-\nu^t}\,q^{-\rho})\\
\eea
where $s_{\lambda/\eta}(\mathbf{x})$ is the skew-Schur function and $\widetilde{Z}_{\nu}(t,q)$ is given by
\bea
\widetilde{Z}_{\nu}(t,q)&=&\prod_{(i,j)\in \nu}\Big(1-q^{a(i,j)}\,t^{\ell(i,j)+1}\Big)^{-1},
\eea
which is related to the Macdonald polynomials
\bea
t^{\frac{\Arrowvert\nu^t\Arrowvert^2}{2}}\,\widetilde{Z}_{\nu}(t,q)&=&P_{\nu}(t^{-\rho};q,t),\\
q^{\frac{\Arrowvert\nu\Arrowvert^2}{2}}\,\widetilde{Z}_{\nu^t}(q,t)&=&P_{\nu^t}(q^{-\rho};t,q).
\eea
For the Macdonald polynomials with this special set of argument we have
\bea
P_{\lambda}(t^{\rho};q,t)=(-1)^{|\lambda|}q^{n(\lambda^{t})}t^{-n(\lambda)}P_{\lambda}(t^{-\rho};q,t).
\eea
The Macdonald polynomials also satisfy
\bea
P_{\lambda}(\mathbf{x};q,t)=P_{\lambda}(\mathbf{x};q^{-1},t^{-1})
\eea
For completeness, let us also review some identities regarding to Schur functions and MacDonald polynomials:
\begin{align}
&s_{\lambda}(\mathbf{x})s_{\mu}(\mathbf{x})=\sum_{\eta}N_{\lambda\mu}^{\eta}s_{\eta}(\mathbf{x})\label{schurfactor}\\
&s_{\lambda/\mu}(\mathbf{x})=\sum_{\eta}N^{\lambda}_{\mu\eta}s_{\eta}(\mathbf{x})\\
&P_{\lambda}(\mathbf{x};q,t)P_{\mu}(\mathbf{x};q,t)=\sum_{\eta}\widehat{N}^{\eta}_{\lambda\mu}P_{\eta}(\mathbf{x};q,t)\label{macdonald}\\
&s_{\eta}(\mathbf{x})=\sum_{\sigma}U_{\eta\sigma}P_{\sigma}(\mathbf{x};q,t)\label{schurmacdonald}
\end{align}
The following relations among Macdonald polynomials prove to be very useful for our computations
\bea\label{ratio}
\frac{P_{\mu}(q^{-\lambda}\,t^{-\rho};q,t)}{P_{\mu}(t^{-\rho};q,t)}&=&\frac{P_{\lambda}(q^{-\mu}\,t^{-\rho};q,t)}{P_{\lambda}(t^{-\rho};q,t)}\\
P_{\sigma}(t^{-\rho},z\,t^{\rho};q,t)&=&P_{\sigma}(t^{-\rho};q,t)\prod_{s\in \sigma}(1-z\,q^{a'(s)}\,t^{-\ell'(s)})\label{product}
\eea
The following sum rules are essential for vertex computations
\begin{align}
&\sum_{\eta}s_{\eta/\lambda}(\mathbf{x})s_{\eta/\mu}(\mathbf{y})=\prod_{i,j=1}^{\infty}(1-x_{i}y_{j})^{-1}\sum_{\tau}s_{\mu/\tau}(\mathbf{x})s_{\lambda/\tau}(\mathbf{y})\\
&\sum_{\eta}s_{\eta^{t}/\lambda}(\mathbf{x})s_{\eta/\mu}(\mathbf{y})=\prod_{i,j=1}^{\infty}(1+x_{i}y_{j})\sum_{\tau}s_{\mu^{t}/\tau}(\mathbf{x})s_{\lambda^{t}/\tau^{t}}(\mathbf{y})\\
&\sum_{\eta}P_{\eta}(\mathbf{x};q,t)P_{\eta^{t}}(\mathbf{x};t,q)=\prod_{i,j=1}^{\infty}(1+x_{i}y_{j})
\end{align}
We have considered the normalized amplitudes, both the open and closed amplitudes are infinite series in the K\"{a}hler parameters, however, their ratio is finite as a result of the following identity
\bea\label{infinite}
\prod_{i,j=1}^{\infty}\frac{1-Q\,q^{-\lambda_{i}+j-1/2}t^{i-1/2}}{1-Q\,q^{j-1/2}t^{i-1/2}}=\prod_{s\in\lambda}\left (1-Q\sqrt{\frac{t}{q}}\, q^{-a'(s)}t^{\ell(s)}\right).
\eea

\bibliography{physics}

\end{document}